%% file: main.tex
\documentclass[reprint,aps,pra,superscriptaddress,nofootinbib,preprintnumbers,longbibliography]{revtex4-1}

\usepackage{comment}
\usepackage{graphics}
\usepackage{bm}
\usepackage{color}
\usepackage{amsmath}
\usepackage{amssymb}
\usepackage{mathrsfs}
\usepackage{graphicx}
\usepackage[caption=false]{subfig}
\usepackage{enumitem}
\usepackage{hyperref} 
\usepackage{verbatim}
\usepackage{mathtools}

\newcommand{\ms}{\hspace{0.05em}}

\begin{document}
\title{Demonstration of a spherical Penning trap for single electrons}

\date{\today}

\author{Zirui Fang}
\email{zirui\_fang@g.harvard.edu}
\affiliation{Department of Physics, Harvard University, Cambridge, Massachusetts 02138, USA}

\author{Xing Fan}
\email{xingfan@g.harvard.edu}
\affiliation{Department of Physics, Harvard University, Cambridge, Massachusetts 02138, USA}
\affiliation{Harvard-MIT Center for Ultracold Atoms, Cambridge, MA 02138, USA}

\begin{abstract}
A spherical Penning trap has well-separated, clean microwave resonances, making it attractive for precision measurements of the electron magnetic moment and for dark-photon and axion searches with trapped electrons.
We demonstrate single-electron trapping in a spherical Penning trap and characterize its microwave resonance structure.
The design, single-electron detection, microwave mode characterization, and advantages of this geometry are presented.
\end{abstract}
\maketitle
\section{Introduction}
Single electrons in Penning traps provide an ideal system for the precise measurement of the electron's magnetic moment ($g$-factor)\cite{ElectronMagneticMoment_Fan_PRL_2023,hanneke2008new,odom2006new,1987DehmeltMagneticMoment} and for dark-photon and axion searches\cite{HighlyExcitedDPAxion_Fan2025,Fan_DarkPhoton_Axion_DarkMatter_2022,Imperial_ElectronDarkPhoton_2026}.
Precision measurement of the $g$-factor provides the most precise test of the quantum electrodynamics (QED) calculations\cite{QED_C8_Lapo,QED_C8_nio,QED_C10_nio,atomsTheoryReview2019,volkov2018numerical,volkov2017new,volkov2019calculating,volkov2024calculation,kitano2024qed,Atoms_Review_2019_g-factor} and determination of the fine-structure constant\cite{PhysRevLett.97.030802,morel2020determination,parker2018measurement}.
To determine the $g$-factor, $g=2\omega_s/\omega_c$, the ratio of the spin frequency $\omega_s$ to the cyclotron frequency $\omega_c$ is measured at better than $10^{-12}$ relative precision. 
In this measurement, one of the largest systematic errors is the dispersive shift of $\omega_c$ from the coupling of the electron to the Penning trap's microwave resonances\cite{brown1985cyclotron,brown1985cyclotronPRL,hanneke2011cavity,fan2022_Thesis}.
This is the so-called cavity-QED systematic effect, shifting the measured $g$-factor as $\delta g/g\simeq-\delta\omega_c/\omega_c$.
A typical Penning trap's size is about $d=3$--5~mm\cite{1987DehmeltMagneticMoment,odom2006new,hanneke2008new,ElectronMagneticMoment_Fan_PRL_2023} and its interior is made of conductors, creating several cavity resonances near the electron's cyclotron frequency $\omega_c\sim2\pi\times100$~GHz.
$\omega_c$ is typically set away from a cavity's resonant mode $\omega_M$ to suppress the shift; in this regime, the shift scales as $    \delta\omega_c\propto d^{-3}(\omega_c^2-\omega_M^2)^{-1}$\cite{brown1985cyclotron}.
Making the trap smaller to increase the mode separation does not solve this problem, as the coupling instead becomes stronger for smaller $d$.
Additionally, the shift does not depend on the $Q$-factor of the resonant modes, so one has to engineer the Penning trap geometry so that the cavity resonances stay as far away as possible and couple to the trapped electron as weakly as possible.

For searches for dark photons and axions, single electrons in a Penning trap work as sensitive electric-field detectors in the milli-electron-volt (meV) range \cite{Fan_DarkPhoton_Axion_DarkMatter_2022,HighlyExcitedDPAxion_Fan2025,Imperial_ElectronDarkPhoton_2026}.
The Penning trap's geometry is also important here for achieving high sensitivity to dark photons and axions.
These particles convert into electromagnetic fields (photons) on a conductor's surface, and the converted photons are emitted perpendicularly from the surface\cite{DishAntennaPrinciple_2013}.
In a spherical trap, since the electron is trapped at the center, the perpendicularly emitted photons are focused at the center of the trap, collectively enhancing the detection efficiency.
As a consequence, compared to traditional cylindrical traps\cite{gabrielse1984cylindrical,tan1989one}, a spherical trap enhances the sensitivity by 1--2 orders of magnitude.

In this paper, we present single-electron trapping in a spherical Penning trap, as well as characterization of its microwave resonances using the internal motion of trapped electrons.
The spherical Penning trap has been proposed in Ref.~\cite{brown1986cyclotronSpherical}, and due to its high symmetry, there are fewer modes than in traditional cylindrical Penning traps\cite{DavidHill}.
For example, transverse-electric (TE) modes do not couple to the trapped electron at the center and only transverse-magnetic (TM) modes couple, creating significantly fewer and weaker modes to characterize for a $g$-factor measurement.
For a spherical trap with the same inner volume as the 2023 cylindrical trap\cite{ElectronMagneticMoment_Fan_PRL_2023}, there are only 3 modes that couple to a centered electron to leading order from 0 to 170~GHz, compared to 28 modes for the cylindrical trap\cite{ElectronMagneticMoment_Fan_PRL_2023,fan2022_Thesis}.
The reduced number of modes will allow detailed study of each resonant mode for the electron's $g$-factor measurement.

The paper is structured as follows. Sec.~\ref{sec:TrapDesign} presents the spherical trap's design as well as single-electron detection. Sec.~\ref{sec:Microwave} presents the measured microwave resonances. Sec.~\ref{sec:Discussion} discusses the gain for the $g$-factor measurement and for dark-photon and axion searches, and Sec.~\ref{sec:Conclusion} summarizes this paper.

\section{Spherical Trap Geometry and Single-Electron Detection}
\label{sec:TrapDesign}
\begin{figure}[]
    \centering
\includegraphics[width=0.97\linewidth]{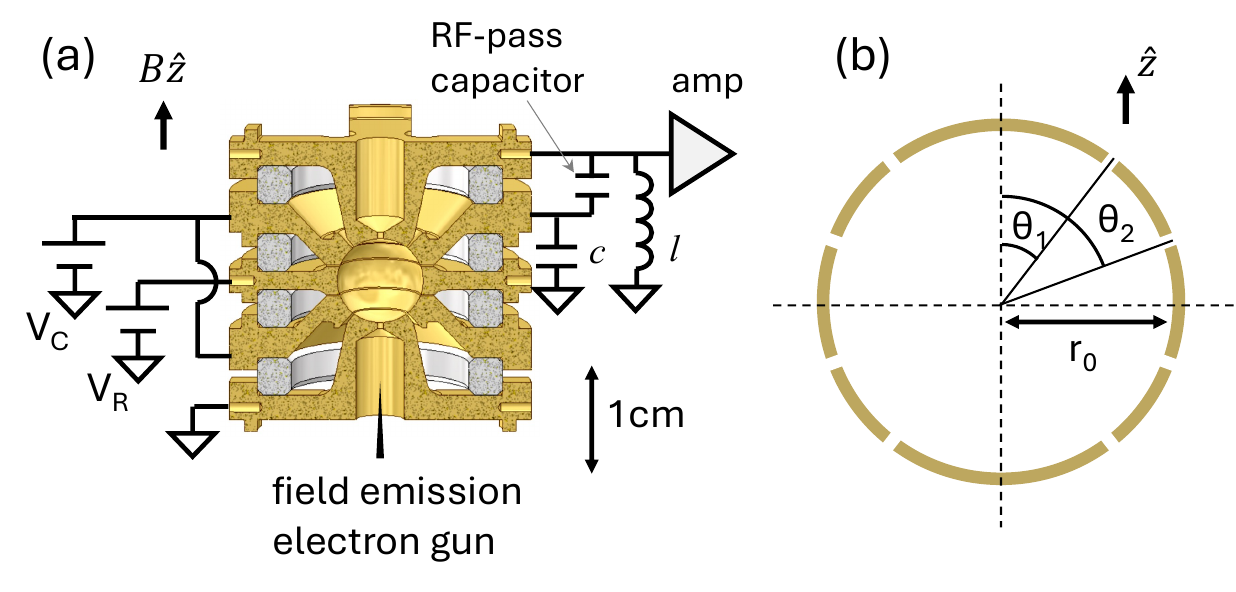} 
     \caption{(a) The spherical trap's geometry and (b) definition of the coordinates of the interior trap volume.}
    \label{fig:Design}
\end{figure}

Figure~\ref{fig:Design} shows the geometry of the spherical trap as well as the definition of coordinates.
The trap consists of five segmented electrodes with an inner spherical radius of $r_0=3.431$~mm (measured in Sec.~\ref{sec:Microwave}).
The electrodes are made from gold-plated copper 101, separated by quartz spacers.
The whole trap is cooled to $T=4.5$~K using a pulse-tube refrigerator (JMTE-Insert, JASTEC, Inc.), and a tunable magnetic field of $B=0$--6 Tesla is applied using a superconducting magnet (JMTD-6T152SS, JASTEC, Inc.).
Electrons are loaded from a field-emission electron gun made of tungsten.
The electric trap potential near the center along the $z$-axis is given by
\begin{subequations}
\begin{align}
\phi(z)=&-V_R\sum_{\substack{k=0 \\\mathrm{even} }}C_{k}\left(\frac{z}{r_0}\right)^k\\
=&-V_R\sum_{\substack{k=0 \\\mathrm{even} }}\left(C^0_k+D_k\frac{V_C}{V_R}\right)\left(\frac{z}{r_0}\right)^k,
\end{align}
\end{subequations}
where $V_R$ is the ring voltage, $V_C$ is the compensation voltage (Fig.~\ref{fig:Design}), and $C^0_k$ and $D_k$ are geometrical parameters given in terms of the Legendre polynomials $P_l(x)$ as
\begin{subequations}
\begin{align}
C^0_k=&P_{k+1}(\cos\theta_2)-P_{k-1}(\cos\theta_2)\nonumber\\
&-P_{k+1}(0)+P_{k-1}(0)\\
D_k=&P_{k+1}(\cos\theta_1)-P_{k-1}(\cos\theta_1)\nonumber\\
&-P_{k+1}(\cos\theta_2)+P_{k-1}(\cos\theta_2)
\end{align}
\end{subequations}
We design the trap with $\theta_1=38.88^\circ$ and $\theta_2=69.56^\circ$ to achieve the orthogonalized condition ($D_2=0$) and the compensated condition ($C_4=C_6=0$ at $V_C/V_R=0.633$) simultaneously\cite{brown1986cyclotronSpherical}.
Under this condition, the anharmonicity ($C_4$ and $C_6$) of the trap potential can be tuned by changing $V_C$ without changing the axial frequency $\omega_z=\sqrt{2C_2eV_R/(m_er_0^2)}$ ($m_e$: electron's mass and $e$: elementary charge).

A helical inductor with $l=200$~nH, made of oxygen-free high-conductivity copper, is attached to the top endcap and compensation electrodes.
These two electrodes are connected by a 5~nF RF-pass capacitor to collect image charge from both of them.
Together with these electrodes' parasitic capacitance to ground, $c=20$~pF, they form a resonator at the axial frequency $\omega_z/(2\pi)=76.8$~MHz with a quality factor $Q=920$.
This resonator damps the electron's axial oscillation with a decay constant of\cite{ElectronCalorimeter}
\begin{equation}
    \gamma=N\gamma_z=N\frac{1}{m_e}\left(\frac{ec_I}{2r_0}\right)^2R,\label{eq:gamma_z}
\end{equation}
where $N$ is the number of trapped electrons, $R=Q\omega_zl$ is the effective parallel resistance of the resonator, and 
\begin{equation}
    c_I=[P_{2}(1)-P_{0}(1)]-[P_{2}(\cos\theta_2)-P_{0}(\cos\theta_2)]
\end{equation}
is the image charge constant.
We achieve $c_I=1.32$ for RF-shorted detection from the top endcap and compensation electrodes, giving $\gamma_z/(2\pi)=14(2)$~Hz, where the uncertainty mainly comes from the estimate of the capacitance.

\begin{figure}[]
    \centering
\includegraphics[width=0.8\linewidth]{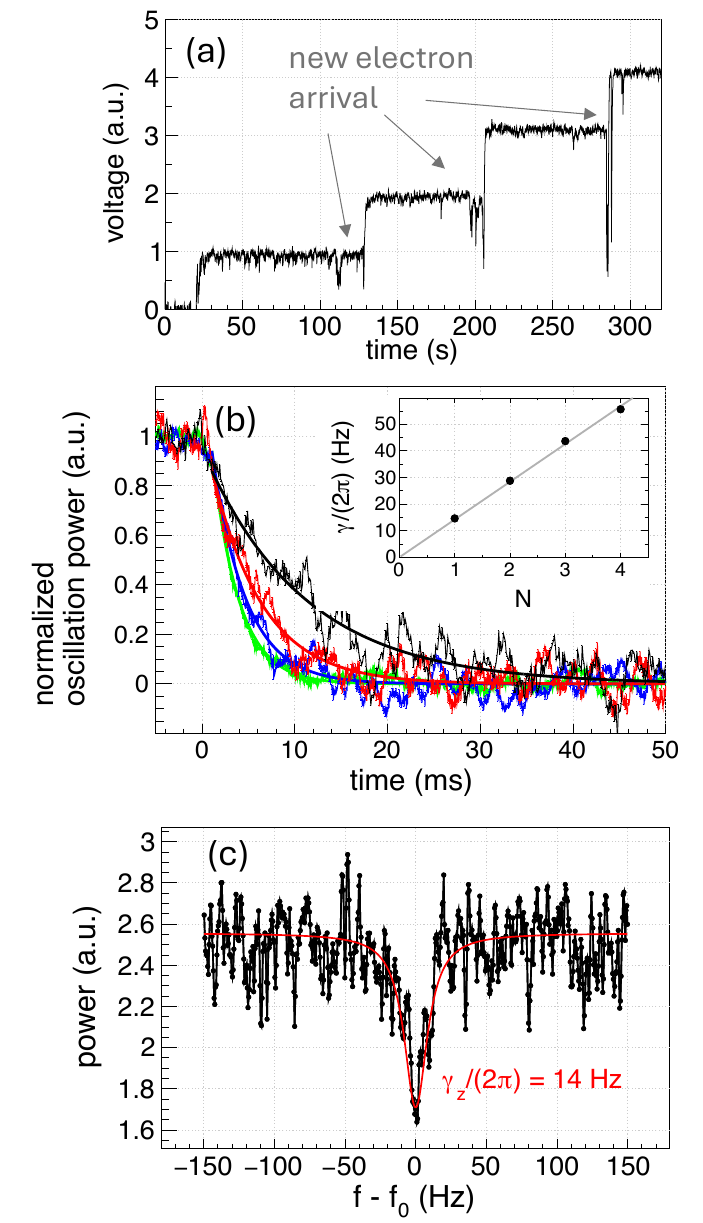} 
     \caption{(a) Counting the number of loaded electrons by monitoring the oscillation amplitude with a parametric drive while firing the electron gun. (b) Direct measurement of $\gamma=N\gamma_z$ by exciting to a large amplitude and monitoring its decay. Different colors correspond to different numbers of trapped electrons. The inset shows the linearity of the measured $\gamma$. (c) Fourier transform of the output signal, showing a dip of a single electron.}
    \label{fig:Dip}
\end{figure}

Fig.~\ref{fig:Dip} shows the counting of the number of trapped electrons. 
In Fig.~\ref{fig:Dip}(a), a strong parametric drive at $\omega_\mathrm{drive}=2\omega_z$ is applied at the bottom electrode and the detected voltage from the detector is monitored, while the field-emission gun is continuously and weakly fired\cite{Tan_Parametric_PRA_1993}.
A stepwise increase in the detected voltage is visible as each new electron arrives.
Fig.~\ref{fig:Dip}(b) shows the direct measurement of $\gamma$ in the time domain.
A strong parametric drive is applied to excite the electron's axial oscillation to about 150~$\mu$m.
After the drive is turned off at $t=0$, the exponential decay is fitted to extract $\gamma$.
Each curve with a different color corresponds to a different number of trapped electrons, loaded by deliberately stopping the field-emission gun after certain steps in Fig.~\ref{fig:Dip}(a).
Fig.~\ref{fig:Dip}(c) shows the Fourier transformation of the trap's output signal without any drive applied.
The electron's signal appears as a ``dip'', and the width corresponds to $\gamma$\cite{ElectronCalorimeter}.
We show the dip for a single electron, demonstrating the detection capability of the trap.

\input{Chapters/CH3/CH3}

\section{Discussion}
\label{sec:Discussion}
Based on the measured microwave spectrum in Sec.~\ref{sec:Microwave}, we discuss the possible improvement for the $g$-factor measurement and for dark-photon and axion searches.
\subsection{The $g$-factor Measurement}
\begin{figure}
    \centering
\includegraphics[width=\linewidth]{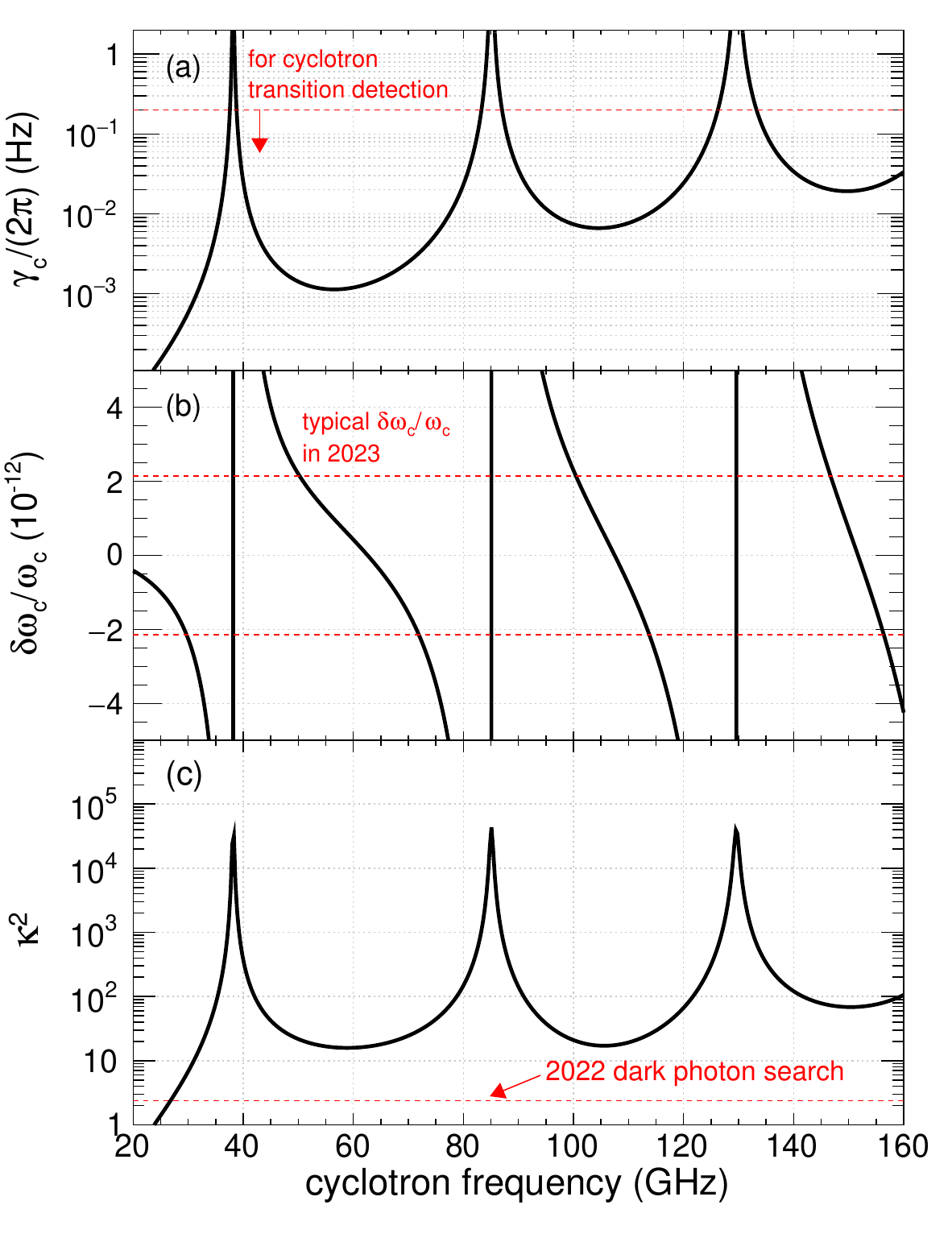} 
     \caption{(a) Calculated cyclotron damping rate $\gamma_c$. The typical condition to detect a quantum cyclotron transition is also shown by the red dotted line. (b) Cavity shift $\delta\omega_c/\omega_c$ from the measured microwave resonances, compared to the typical shift in the 2023 measurement\cite{ElectronMagneticMoment_Fan_PRL_2023}. (c) Enhancement factor $\kappa^2$ for dark-photon and axion searches from the measured spherical trap's microwave resonances.}
    \label{fig:Gain}
\end{figure}

The exact calculation of the cyclotron damping rate $\gamma_c$ and the correction of the cyclotron frequency $\delta\omega_c$ is given in Ref.~\cite{brown1985cyclotron,brown1985cyclotronPRL,brown1986cyclotronSpherical}.
For a specific cavity mode $M$, the single-mode approximation gives an instructive picture:
\begin{eqnarray}
\gamma_c&\simeq&\frac{2\lambda_M^2\omega_c^2\Gamma_M}{(\omega_c^2-\omega_M^2)^2+\omega_c^2\Gamma_M^2}  \label{eq:gamma_cSingleMode}\\
    \frac{\delta\omega_c}{\omega_c}&\simeq&\frac{\lambda_M^2}{\omega_c^2-\omega_M^2},
    \label{eq:lambdaM2}
\end{eqnarray}
where $\Gamma_M=\omega_M/Q_M$ is the full width at half maximum, and $\lambda_M^2$ is the coupling constant for the mode\cite{brown1985cyclotron}:
\begin{eqnarray}    
    \lambda_M^2&=&\frac{e^2}{m_e \epsilon_0}\frac{\left|E_{M}(\mathbf r)_x\right|^2+\left|E_{M}(\mathbf r)_y\right|^2}{\displaystyle \int_V \left|\mathbf E_M(\mathbf{r'})\right|^2 \, d^3 \mathbf{r'}},
    \label{eq:lambdaM2Def}
\end{eqnarray}
where $\epsilon_0$ is the vacuum permittivity, the integral is for the whole cavity volume $V$, and $\mathbf{r}$ denotes the electron's location, usually near the center.
Importantly, $\gamma_c$ depends on the mode's $Q$-factor, but $\delta\omega_c$ does not; $\delta\omega_c$ is purely determined by the coupling and the separation of the modes.
For the spherical trap, the calculated $\gamma_c$ and $\delta\omega_c$ are shown in Fig.~\ref{fig:Gain}(a) and (b).

To detect the quantum cyclotron transition, $\gamma_c/(2\pi)<0.2$~Hz is typically required [red dashed line in Fig.~\ref{fig:Gain}(a)]\cite{Durso_SelfExcitation_2005}.
Due to the large separation of the modes, a broad range of suitable choices for $\omega_c$ is available.
Fig.~\ref{fig:Gain}(b) compares the correction $\delta\omega_c/\omega_c$ with the average of the corrections in the 2023 $g$-factor measurement\cite{ElectronMagneticMoment_Fan_PRL_2023}, showing the widely available range with smaller correction $\delta\omega_c$.
The uncertainty of $\lambda_M^2$ is the largest source of systematic error.
In the 2023 $g$-factor measurement~\cite{ElectronMagneticMoment_Fan_PRL_2023}, $\lambda_M^2$ was only estimated from the cylindrical trap's imperfection and not directly measured due to the dense mode structure.
With fewer modes in the spherical trap, a new strategy becomes possible: directly measuring $\lambda_M^2$ for each individual mode.
One can set $\omega_c$ near each of the cavity resonance and measure the linewidth of the anomaly transition to determine $\gamma_c$ precisely.
We expect that this could allow the first direct measurement of $\lambda_M^2$ at the 0.5\% level, which will then be limited by the frequency calibration and the determination of $Q$, significantly reducing the uncertainty of $\delta\omega_c$.
The largest systematic error with this strategy will then become the uncertainty of $\omega_M$'s center frequency, and the $g$-factor's systematic error could be reduced to $\delta g/g\simeq2$--$3\times10^{-14}$\cite{fan2022_Thesis,QLS_Electron_2025}.
\subsection{Dark-Photon and Axion Searches}
The electron's motion in a Penning trap serves as a sensitive meV electric-field detector for dark photons and axions\cite{Fan_DarkPhoton_Axion_DarkMatter_2022,HighlyExcitedDPAxion_Fan2025,Imperial_ElectronDarkPhoton_2026}.
Dark photons and axions convert on the trap's surface and are emitted perpendicularly from the wall\cite{DishAntennaPrinciple_2013}.
A spherical trap provides two advantages: First, the simple mode structure allows searches for dark photons and axions in a broad range without the need to avoid cavity resonances, as shown in Fig.~\ref{fig:Gain}(a).
Second, the converted photons from the entire surface are focused at the center of the trap, giving an enhancement factor $\kappa^2$ approximately $\kappa^2\simeq(r_0\omega_c/c)^2$.
The exact enhancement factor is given in Ref.~\cite{HighlyExcitedDPAxion_Fan2025}, and we cite only the result:
\begin{widetext}
\begin{eqnarray}
\kappa^2 &=&  \left| \sum_{p=1}^{\infty} \frac{4}{3}\frac{\omega^2}{\left(\frac{u_{1p}'c}{r_0}\right)^2-\omega^2}\frac{  -{u_{1p}'}^5  j_1(u_{1p}')}{ {u_{1p}'}^4+ (2 u_{1p}' -\frac{1}{2}{u_{1p}'}^3)\sin\left(2 u_{1p}'\right) -\left[1+\cos(2 u_{1p}')\right]{u_{1p}'}^2-1 +\cos(2 u_{1p}') }\right|^2.
\end{eqnarray}
\end{widetext}
Fig.~\ref{fig:Gain}(c) shows the calculated $\kappa^2$ for the demonstrated spherical trap as well as the enhancement factor used in the 2022 dark-photon search\cite{Fan_DarkPhoton_Axion_DarkMatter_2022}.
Over a wide frequency range, $\kappa^2$ is larger by 2--3 orders of magnitude in the spherical trap.
Due to the scaling $\kappa^2\propto(r_0\omega_c/c)^2$, a larger spherical trap would provide an even further enhancement, but a larger trap will also reduce the detection sensitivity, $\gamma_z\propto {r_0^{-2}}$ [Eq.~\eqref{eq:gamma_z}].
Determining how large the trap can be made is important for dark-photon and axion searches and will be a future research topic.
\section{Conclusion}
\label{sec:Conclusion}
We have developed and demonstrated single-electron trapping and detection in a spherical Penning trap. 
We have also characterized the trap's microwave resonances and observed a sparse mode structure, as expected from the high symmetry of the spherical geometry.
The reduced number of modes suggests that the spherical trap could be a promising geometry for future precision electron $g$-factor measurements as well as dark-photon and axion searches.

\begin{acknowledgments}
This work is supported by the National Science Foundation Award No.~2317134. X.F. acknowledges support from the Masason Foundation.
\end{acknowledgments}

\section*{Data Availability}

The data that support the findings of this article are available from the corresponding author upon reasonable request.

\bibliography{PenningTrapExperimentRefs}

\end{document}

%% file: Chapters/CH3/CH3.tex
\section{Microwave Cavity Modes}
\label{sec:Microwave}

The microwave resonances of the spherical trap are characterized using the cyclotron-cooling response of a cloud of electrons.
The data presented here were taken with $N\simeq 300$ electrons, but we have checked that the result does not qualitatively depend on $N$ for $N=100$--600.
A strong parametric drive at $\omega_{\rm drive}\simeq 2\omega_z$ is applied to the bottom electrode to heat the internal motion of the
cloud, and the steady-state amplitude of the axial signal is recorded as the magnetic field $B$ is swept\cite{Tan_Parametric_PRA_1993,Tan_Parametric_Synchronization_PRL_1991}.
Empirically, the response increases as the cloud's internal temperature cools\cite{Dehmelt_MicrowaveCavity_Obseration_1987}, and the internal temperature is set by the cyclotron damping rate $\gamma_c$ in the surrounding cavity\cite{Tan_Parametric_Synchronization_PRL_1991}.
Sweeping the cyclotron frequency $\omega_c=eB/m_e$ therefore traces out the entire microwave coupling rate of the trap.

For a perfectly conducting spherical boundary of radius $r_0$, the
source-free Maxwell equations allow two sets of eigenmodes, TE and TM\cite{DavidHill}.
Their angular dependence is given by the spherical harmonics $Y_n^m(\theta,\varphi)$ and their radial dependence by the spherical Bessel function $j_n(x)$.
Requiring the tangential electric field to vanish on $r=r_0$ gives the eigenvalue conditions for TE and TM modes\cite{DavidHill}
\begin{subequations}\label{eq:cavity_modes}
\begin{align}
    j_n\left(\frac{\omega_{mnp}^\mathrm{TE}}{c}r_0\right)&=0,\\[2pt]
    \left.\frac{d}{dr}\left[r\,j_n\left(\frac{\omega_{mnp}^\mathrm{TM}}{c} r\right)\right]\right|_{r=r_0}&=0.
\end{align}
\end{subequations}
Using $u_{np}$ as the $p$-th
zero of $j_n(x)$, and $u'_{np}$ as the $p$-th zero of
$\frac{d}{dx}\left[xj_n(x)\right]=j_n(x)+x\,j_n'(x)$, the corresponding eigenfrequencies are
\begin{align}
\omega^\mathrm{TE}_{mnp}=\frac{cu_{np}}{r_0},
\qquad
\omega^\mathrm{TM}_{mnp}=\frac{cu'_{np}}{r_0},
\label{eq:cavity_freqs}
\end{align}
with each $(n,p)$ being $(2n+1)$-fold degenerate in $m$ for a perfect sphere. Both TE and TM modes start at $n=1$.

A given mode's coupling to the cyclotron motion of a single electron is set by the value of the mode's transverse electric field at the electron's location\cite{brown1985cyclotron,brown1986cyclotronSpherical}.
At $r=0$, TE modes have zero radial electric field, $E_r\equiv 0$, and their transverse components ($E_\theta$, $E_\phi$) scale as $r^n$ near the origin, so every TE mode decouples from a centered electron.
TM modes have $E_r\propto r^{n-1}$ at small $r$; only the $n=1$
manifold has a non-vanishing field at the origin.
Within TM, $n=1$, the electric field of the $m=0$ component is linearly polarized along
$\hat z$ and does not couple to the cyclotron motion, while the $m=\pm 1$ components are polarized in the $xy$-plane and couple to the cyclotron motion.
The cyclotron motion of a centered electron therefore couples, to leading order, only to $\mathrm{TM}_{m=\pm1\,n=1\,p}$, and in the typically experimentally accessible window, $0<\omega_c/(2\pi)<168$~GHz ($0~\textrm{T}<B<6~\textrm{T}$)\cite{van1999ultrastable,fan2019gaseous,RevModPhys.58.233}, only three such modes
exist: $\mathrm{TM}_{\pm1\ms1\ms1}$, $\mathrm{TM}_{\pm1\ms1\,2}$, and $\mathrm{TM}_{\pm1\ms1\,3}$.

\begin{figure*}[t]
    \centering
    \includegraphics[width=0.9\linewidth]{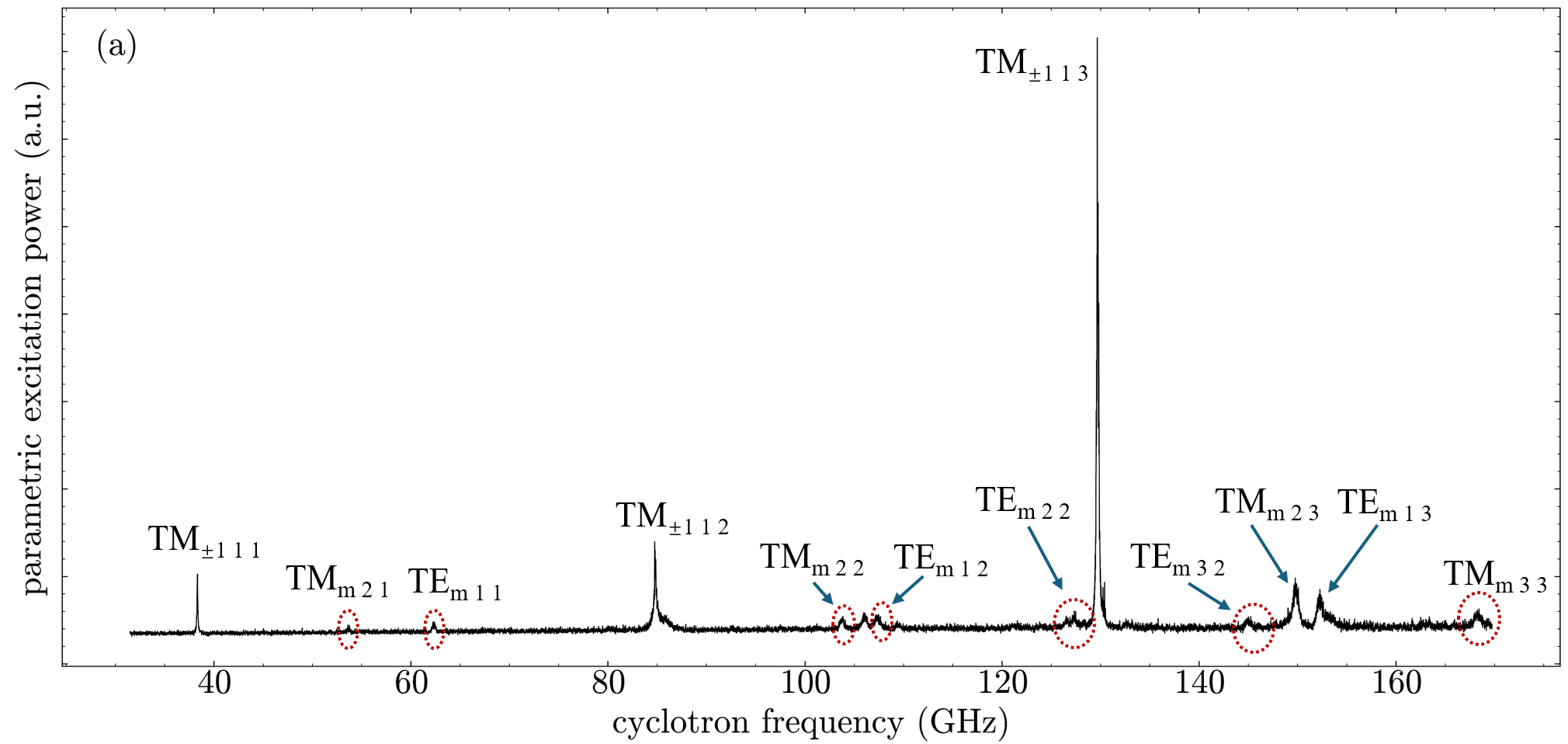}\\[4pt]
    \includegraphics[width=0.33\linewidth]{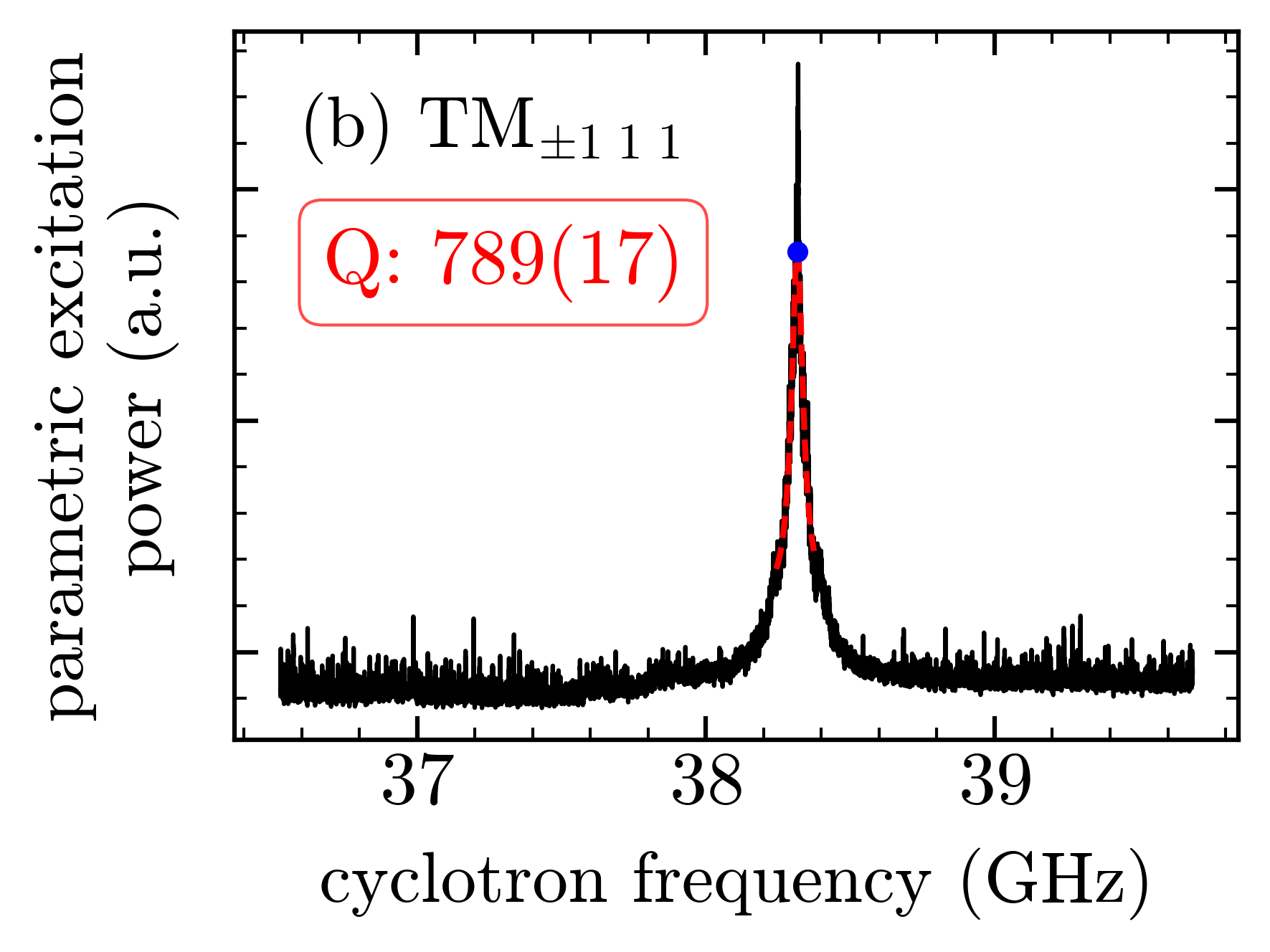}\hfill
    \includegraphics[width=0.33\linewidth]{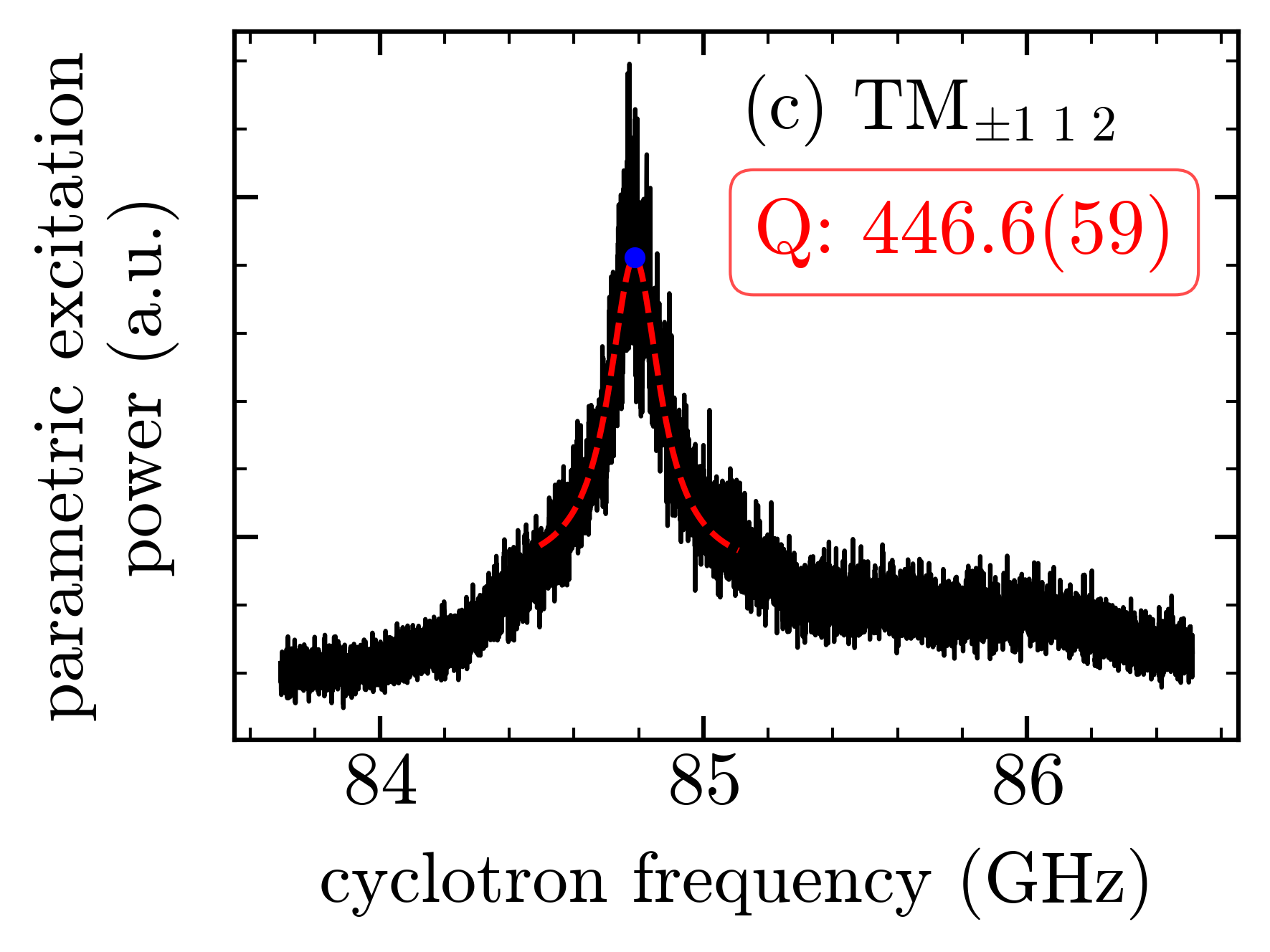}\hfill
    \includegraphics[width=0.33\linewidth]{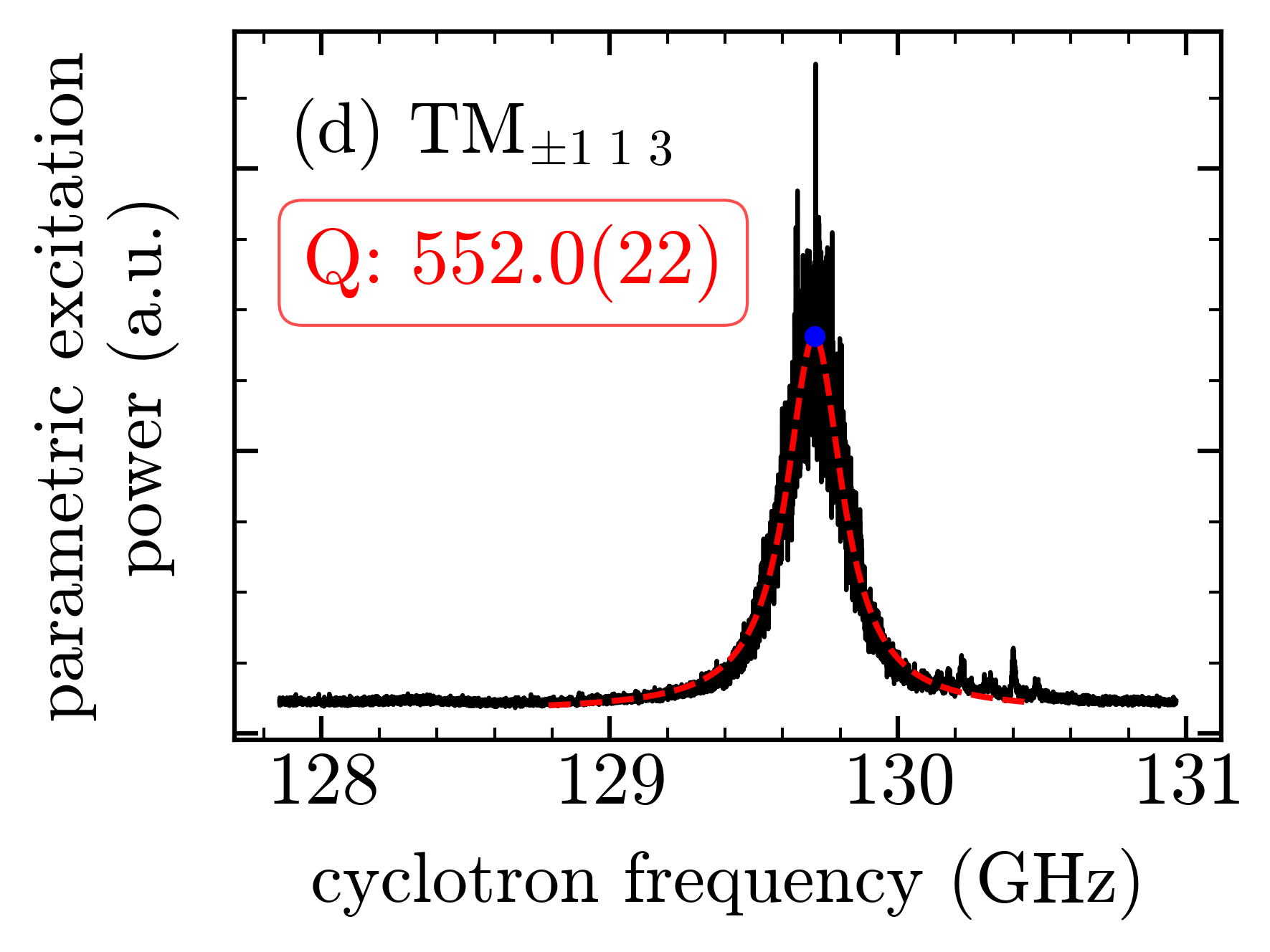}
    \caption{Measured microwave spectrum of the spherical trap.
    (a) Fast magnet-current sweep over $B=1.12$--$6.07$~T
    ($f_c=31.40$--$170.02$~GHz).
    The three dominant peaks at $38$, $85$, and $130$~GHz
    are identified with $\mathrm{TM}_{\pm1\ms1\ms1}$, $\mathrm{TM}_{\pm1\ms1\,2}$, and
    $\mathrm{TM}_{\pm1\ms1\,3}$, respectively. The additional, weaker peaks
    labeled in (a) correspond to the higher-order modes
    listed in Table~\ref{tab:HigherModes}.
    (b)--(d) Slow re-scans of the three $\mathrm{TM}_{\pm1\ms1\ms p}$ resonances at $\sim 0.157$~GHz/min, together with the Lorentzian fits (red dashed) used to extract the center frequency and quality factor quoted in each panel.}
    \label{fig:Microwave}
\end{figure*}

We map the cavity spectrum in two stages.
First, the superconducting magnet's field is swept over $B=1.12$--$6.07$~T, corresponding to $\omega_c/(2\pi)=31.40$--$170.02$~GHz, at 0.217~T/min (6.1~GHz/min), while the integrated cloud
response is recorded; the result is shown in
Fig.~\ref{fig:Microwave}(a).
Three dominant peaks are immediately apparent near $38$, $85$, and
$130$~GHz, in excellent qualitative agreement with the predicted
$\mathrm{TM}_{\pm1\ms1\,p}$ ($p=1,2,3$) structure.
Second, each of the three modes is re-scanned at a much slower rate of 0.0056~T/min ($0.157$~GHz/min).
For both fast and slow scans, we sweep the magnetic field up and down to cancel hysteresis from the ramping process.
The resulting traces
[Fig.~\ref{fig:Microwave}(b)--(d)] are fitted to a Lorentzian
\begin{align}
S(f)=A_M\frac{\left(\frac{1}{2}\frac{\Gamma_M}{2\pi}\right)^2}{(f-f_M)^2+\left(\frac{1}{2}\frac{\Gamma_M}{2\pi}\right)^2}+C,
\end{align}
with center frequency $f_M=\omega_M/(2\pi)$, full width
$\Gamma_M=\omega_M/Q_M$, and a constant background $C$.
The center frequencies extracted in this way have two uncertainty contributions: the statistical fit uncertainty
($0.19$, $0.53$, and $0.35$~MHz for the three modes), and the absolute calibration of the superconducting magnet.
The magnetic field is calibrated by measuring the magnetron frequency at the $\delta B/B=7.05\times 10^{-4}$ level.
Because the cyclotron-frequency axis of Fig.~\ref{fig:Microwave} is
set by the magnet, the latter propagates directly into the extracted
mode centers as $\delta f_M = (\delta B/B)\,f_M$, and dominates the
fit uncertainty by two to three orders of magnitude.
Combining the two contributions in quadrature, the centers and loaded
quality factors are
\begin{align}
f_{\mathrm{TM}_{\pm1\ms1\ms1}}&=38.317(27)~\mathrm{GHz},\ \ Q=789(17),\nonumber\\
f_{\mathrm{TM}_{\pm1\ms1\,2}}&=84.787(60)~\mathrm{GHz},\ \ Q=446.6(59),\\
f_{\mathrm{TM}_{\pm1\ms1\,3}}&=129.710(91)~\mathrm{GHz},\ \ Q=552.0(22).\nonumber
\end{align}

A simultaneous least-squares fit of these three centers to
$f_{\mathrm{TM}_{\pm1\ms1\,p}}=(c/2\pi r_0)\,u'_{1p}$, with $r_0$ as the single free
parameter, yields a fit residual on $r_0$ of $5~\mu$m.
The systematic contribution from the magnet calibration is
$\delta B/B\times r_0\approx 2.4~\mu$m, which is fully correlated
across the three modes and therefore enters only as a rescaling of $r_0$.
Combining the two contributions in quadrature gives
\begin{align}
r_0=3.431(6)~\mathrm{mm},
\end{align}
with the total uncertainty now dominated by the fit residual rather
than the magnet calibration.

The fitted cold-cavity radius is $125~\mu$m smaller than the designed room-temperature dimension of $r_0=3.556$~mm. Thermal contraction of
the gold-plated copper electrodes between room temperature and 4.5~K is
only $\sim 0.3\%$ of $r_0$, i.e.\ $\sim 11~\mu$m, and therefore accounts
for less than 10\% of this 125~$\mu$m difference. The remainder is dominated by small geometric corrections from the segmented-electrode gaps, which shift the effective electromagnetic boundary away from the ideal sphere assumed in Eq.~\eqref{eq:cavity_modes}, together with the machining tolerance ($25~\mu$m for each electrode).
Table~\ref{tab:MainModes} summarizes the three coupled resonances
together with their theoretical frequencies at the fitted $r_0$, showing good agreement.
\begin{table}[t]
\centering
\begin{tabular}{c|c|c|c|c}
\begin{tabular}{@{}c@{}}resonant\\mode\end{tabular} &
\begin{tabular}{@{}c@{}}observed\\frequency (GHz)\end{tabular} &
\begin{tabular}{@{}c@{}}calculated\\frequency (GHz)\end{tabular} &
measured $Q$ \\
\hline
$\mathrm{TM}_{\pm1\ms1\,1}$ & $38.317(27)$   & $38.16$  & $789(17)$    \\
$\mathrm{TM}_{\pm1\ms1\,2}$ & $84.787(60)$   & $85.07$  & $446.6(59)$  \\
$\mathrm{TM}_{\pm1\ms1\,3}$ & $129.710(91)$  & $129.57$ & $552.0(22)$  
\end{tabular}
\caption{Microwave resonances strongly coupled to the centered cyclotron motion. Observed frequencies and $Q$ factors are extracted from Lorentzian fits to the slow scans in Fig.~\ref{fig:Microwave}(b)--(d). The quoted frequency uncertainty is the quadrature sum of the Lorentzian-fit uncertainty and the magnet-calibration contribution $\delta B/B\times f$, with $\delta B/B=7.05\times 10^{-4}$; the latter dominates. The calculated frequencies are evaluated using the best-fit value $r_0=3.431(6)~\mathrm{mm}$~\cite{brown1986cyclotronSpherical}.}
\label{tab:MainModes}
\end{table}

In addition to the three $\mathrm{TM}_{\pm1\ms1\,p}$ resonances, a number of weaker peaks are visible in Fig.~\ref{fig:Microwave}(a).
We identify these with $n\ge 2$ TM modes and $n\ge 1$ TE modes, which couple to the cyclotron motion when the cloud is displaced from the geometric center of the sphere.
For a displacement $\Delta r$ (either radially or axially), the coupling to $\mathrm{TM}_{mnp}$ ($n\ge 2$) is suppressed by $(\Delta r/r_0)^{2(n-1)}$ relative to the on-center $\mathrm{TM}_{\pm1\ms1\,p}$ coupling, and the coupling to $\mathrm{TE}_{mnp}$ is suppressed by $(\Delta r/r_0)^{2n}$ analogously.
The observed amplitudes of these higher-order peaks therefore reflect the displacement of the electron cloud from the center.

To estimate the displacement, the integrated area gives the ratio of the coupling rate of $\mathrm{TM}_{m\ms2\ms2}$ and $\mathrm{TM}_{\pm1\ms1\,2}$, $\lambda^2_{\text{TM}_{m\ms2\ms2}}/\lambda^2_{\text{TM}_{\pm1\ms1\ms2}}\simeq 1/8.27$ .
The displacement is mainly caused by the parametric drive, driving the axial oscillation strongly, so we approximate $\Delta r\simeq\Delta z$.
Under this approximation, the leading coupling of TM$_{m22}$ is from the $m=\pm1$ modes.
The coupling rates of these modes are given by Eq.~\eqref{eq:lambdaM2Def}, and numerically,
\begin{equation}
    \frac{\lambda^2_{\text{TM}_{\pm1\ms2\ms2}}}{\lambda^2_{\text{TM}_{\pm1\ms1\ms2}}}=\frac{4.0\times10^{12}~\text{s}^{-2}~\left(\frac{\Delta z}{r_0}\right)^2}{3.2\times10^{11}~\text{s}^{-2}}=\frac{1}{8.27},
\end{equation}
giving $\Delta z\simeq 0.34$~mm.
This estimate is qualitatively consistent with the estimate from the parametric excitation amplitude and the amplitude in thermal equilibrium $\sqrt\frac{k_BT}{m_e\omega_z^2}\simeq15$~$\mu$m, although the cloud has a large internal motion, so the comparison is valid only as an order-of-magnitude estimate.

Table~\ref{tab:HigherModes} additionally lists the nine higher-order modes identified in Fig.~\ref{fig:Microwave}(a) together with their theoretical frequencies at the fitted $r_0$; the quoted uncertainty on each observed frequency is the magnet-calibration contribution $\delta B/B\times f$, which dominates over the fit uncertainty for these comparatively weak peaks.
All other modes predicted by Eq.~\eqref{eq:cavity_modes} below
170~GHz lie below the noise floor of the cloud-cooling measurement,
consistent with their predicted $(\Delta r/r_0)^{2(n-1)}$ or higher
suppression.

\begin{table}[t]
\centering
\begin{tabular}{c|c|c}
\begin{tabular}{@{}c@{}}resonant\\mode\end{tabular} &
\begin{tabular}{@{}c@{}}observed\\frequency (GHz)\end{tabular} &
\begin{tabular}{@{}c@{}}calculated\\frequency (GHz)\end{tabular}\\
\hline
$\mathrm{TM}_{m 2 1}$ & $53.593(38)$   & $53.83$  \\
$\mathrm{TE}_{m 1 1}$ & $62.246(44)$   & $62.49$  \\
$\mathrm{TM}_{m 2 2}$ & $103.850(73)$  & $103.52$ \\
$\mathrm{TE}_{m 1 2}$ & $107.438(76)$  & $107.44$ \\
$\mathrm{TE}_{m 2 2}$ & $127.214(90)$  & $126.49$ \\
$\mathrm{TE}_{m 3 2}$ & $145.247(102)$ & $144.88$ \\
$\mathrm{TM}_{m 2 3}$ & $149.782(106)$ & $148.99$ \\
$\mathrm{TE}_{m 1 3}$ & $152.273(107)$ & $151.65$ \\
$\mathrm{TM}_{m 3 3}$ & $168.301(119)$ & $167.78$ 
\end{tabular}
\caption{Higher-order microwave resonances identified in
Fig.~\ref{fig:Microwave}(a). These modes couple to the cyclotron
motion only at finite displacement $\Delta r$ from the trap center.
The quoted frequency uncertainty is the magnet-calibration
contribution $\delta B/B\times f$ alone, with
$\delta B/B=7.05\times 10^{-4}$.
The calculated frequencies are evaluated at the fitted $r_0=3.431(6)~\mathrm{mm}$.}
\label{tab:HigherModes}
\end{table}